\documentclass[a4paper,11pt]{article}
\pdfoutput=1 

\usepackage{jinstpub} 
\usepackage{amsfonts}
\usepackage{booktabs}
\usepackage{lineno}
\newcommand{\pbfd}{PbF$_2$ }

\usepackage[numbers,sort&compress]{natbib}
\begin{document}
\title{\boldmath Radiation study of Lead Fluoride crystals}

\author[a]{A. Cemmi,}
\author[b]{A. Colangeli,}
\author[a]{B. D'orsi,}
\author[a]{I. Di Sarcina,}
\author[c]{E. Diociaiuti,}
\author[b]{S. Fiore,}
\author[d]{D. Paesani,}
\author[b]{M. Pillon,}
\author[c]{A. Saputi,}
\author[c]{I. Sarra,}
\author[e]{D. Tagnani}

\affiliation[a]{ENEA R.C. Casaccia, TER,\\ Via Anguillarese 301,00123 Rome, Italy}
\affiliation[b]{ENEA, Department of Fusion and Technology for Nuclear Safety and Security,\\ Via Enrico Fermi45,00054 Frascati Italy}
\affiliation[c]{National  Laboratories of Frascati,\\Via Enrico Fermi 54, 00054 Frascati, Italy}
\affiliation[d]{Istituto Nazionale di Fisica Nucleare, Sezione di Bari,\\ Via Orabona 4, 70126 Bari, Italy}
\affiliation[e]{Roma Tre University,\\ Via della Vasca Navale 84, 00146 Roma, Italy}

\emailAdd{crilin-contacts@lists.infn.it}

\abstract{ 
Lead fluoride (PbF$_2$) crystals represent an excellent and relatively innovative choice for high resolution electromagnetic calorimeters with high granularity and fast timing for high intensity environments. For this reason two \pbfd crystals, sized $5\times 5 \times 40 $ mm$^3$, were irradiated with $^{60}$Co photons up to $\sim 4$~Mrad and with 14 MeV neutrons up to a $10^{13}$~ n/cm$^2$ total fluence. 
Their loss in transmittance was evaluated at different steps of the photon and neutron irradiation campaign, resulting in  a total of 30\% loss above 350 nm. With crystals always kept in dark conditions, natural and thermal annealing, as well as optical bleaching with 400 nm light, were performed on the irradiated specimens, resulting in a partial recovery of their original optical characteristics.
}

\maketitle
\flushbottom

\section{Introduction}

Lead Fluoride crystals represent a very interesting choice for new generation calorimeters. Differently from the most common crystals used in HEP, \pbfd  is relatively cheap to produce and - being a Cherenkov radiatior - its timing performance is not limited by the typical scintillator  rise and decay times. \pbfd shows a very good transmittance in the UV region with a spectral cut-off edge at $\sim$ ~240 nm. 
Its optical transmittance is a key parameter, since the number of Cherenkov photons  peaks at low wavelenghts as follows \cite{jelly1958cherenkov}:
\begin{equation*}
    dN \propto d\lambda/\lambda^2,
\end{equation*}
a good transmittance at shorter wavelengths allows a better yield of Cherenkov light \cite{ren2001optical}.
A comparison of its properties with respect to other high-density crystals frequently used in electromagnetic calorimetry is reported in Table~\ref{tab:properties} \cite{kozma2002radiation}.
\begin{table}[h!]
    \centering
   \begin{tabular}{cccccc} \toprule
   Crystal & \pbfd & BGO & BaF$_2$ & CsI & CeF$_2$ \\ \midrule
   Density [g/cm$^3$] & 7.7 & 7.13 & 4.87 & 4.51 & 6.16 \\
   Radiation lenght [cm] & 0.93 & 1.1 & 2.1 & 1.9 &1.7 \\
   Moli\`ere radius [cm] & 2.2 & 2.7 & 4.4 & 3.8 & 2.6 \\
   Decay constant [ns] & Ch & 300 & 0.6, 620 & 26 & 5, 30\\  \bottomrule
   \end{tabular}
    \caption{Properties of \pbfd compared to those of other high density crystals}
    \label{tab:properties}
\end{table}

The individual and combined effect of Total Ionizing Dose (TID) and neutrons was evaluated by measuring the resulting deterioration in transmittance for two crystals sized $5\times 5 \times 40 $ mm$^3$, manufactured by SICCAS\cite{SIC} using a melt growth process, thus resulting in a cubic form ~($\beta$-PbF$_2$). 
The first crystal was tested without any kind of wrapping (in the following referred as the ``naked'' one), the other was wrapped with a 100~$\mu$m thick Mylar foil.

\section{Optical properties before the irradiation campaign}
Transmittance measurements are useful to investigate properties and quality of these type of crystals. 
Transmittance is defined as the ratio between the intensity of a light beam attenuated by its passage trough the crystal and its original intensity:
 $ T=\frac{I_0\left(1-R\right)^2e^{-\alpha d}}{I_0}$.\\
The transmittance was measured longitudinally - \textit{i.e.} through the crystal axis - with a PerkinElmer Lambda 950 UV/VIS dual-beam spectrometer \cite{PerkinElmer}. In this case the optical transmittance has been evaluated as follows:
\begin{equation*}
    T=\frac{\frac{S-D}{Ref-D}}{\frac{S_0-D_0}{Ref_0-D_0}},
\end{equation*}
where $S$, $D$ and $Ref$ are respectively the measured, reference and dark signals, while the subscript 0 refers to the baseline measurement performed without the crystal inside the spectrometer.\\

\begin{figure}[h!]
    \centering
     \includegraphics[width=\textwidth]{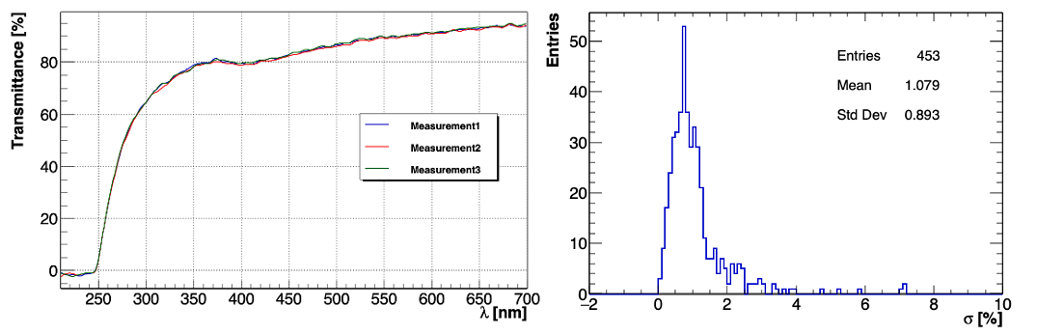}
    \caption{Longitudinal transmittance measurement on ``naked'' crystal (Left). $\sigma$ values for repeated longitudinal transmittance measurements performed removing and re-placing the crystal on the holder (Right).}
    \label{fig:trasm_inizio}
 \end{figure}

Before starting the measurements, a reproducibility test was performed by removing and placing the crystal on the sample holder for each acquisition. The obtained spectra are reported in \figurename~\ref{fig:trasm_inizio} (Left). The spread at each wavelength can be evaluated as:
\begin{equation*}
    \sigma=\frac{T_{max}-T_{min}}{T_{max}}
\end{equation*}
Where $T_{max}$ and $T_{min}$ are respectively the maximum and minimum transmittance values relative to a single set of measurements. 
As shown in \figurename~\ref{fig:trasm_inizio} (Right), a $\sigma$= 1.1\% is associated with the longitudinal transmittance measurement.

\begin{figure}[h!]
     \centering
     \includegraphics[width=.6\textwidth]{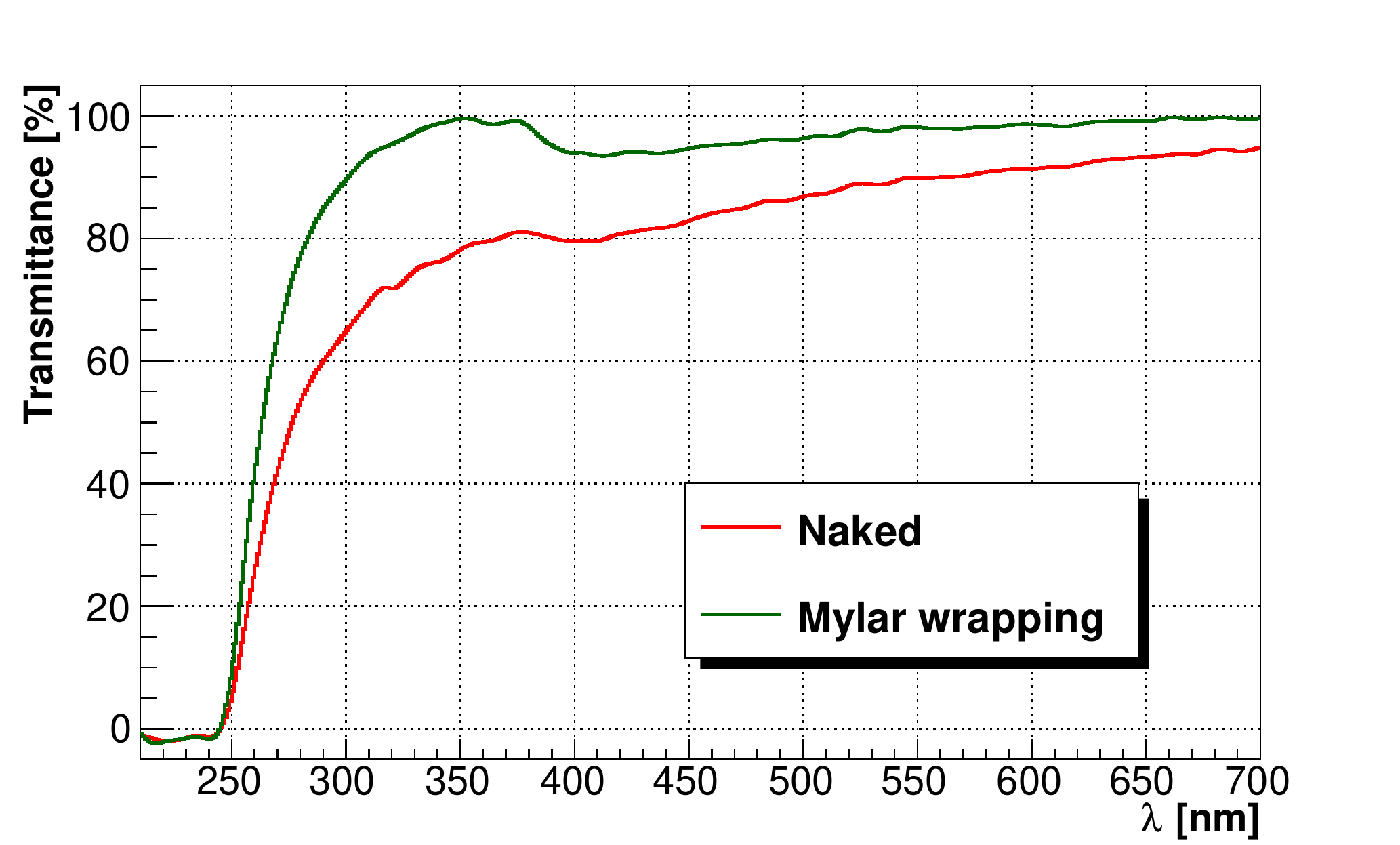}
     \caption{Longitudinal transmittance of the crystals before irradiation.}
     \label{fig:trasmIniz_confronto}
 \end{figure}

 A comparison between the transmittance spectra obtained with the ``naked'' and the Mylar-wrapped crystals is shown in \figurename~\ref{fig:trasmIniz_confronto}. The spectrum of the ``naked'' crystals shows a lower transmittance with respect to the one wrapped with Mylar because of the reflection losses.

\section{Optical properties after $^{60}$Co photons irradiation}
After the reference transmittance measurements, the irradiation phase was carried out at Calliope\cite{Calliope}, a pool-type gamma irradiation facility equipped with a $^{60}$Co radio-isotopic source array housed in a large volume $(7.0\times 6.0 \times 3.9$ m$^3$) shielded cell. The source rack is composed of 25 $^{60}$Co source rods (with  $41 \times 90$ cm$^2$ active area) arranged in a planar geometry, producing photons with $E_{\gamma} = 1.25$ MeV. At the time when the measurements were conducted, the activity of the plant was $1.97\times 10^{15}$ Bq.\\
The crystals were positioned 70 cm away from the source, with their longitudinal axis perpendicular to it, yielding a 100 krad/h dose rate.\\
The transmittance measurements were performed at different irradiation steps. The irradiation campaign  took place over the span of three days for a total absorbed dose of 4.4 Mrad.\\
In Table~\ref{tab:Irrstep} the dose, expressed in krad (air), absorbed in each irradiation step are reported.

\begin{table}[h!]
    \centering
    \begin{tabular}{cc} \toprule
    Irradiation Step & Dose in air [krad] \ \\ \midrule
   I & 30.2 \\
   II & 89.88\\
   III & 2082 \\
   IV & 4031.8 \\
   V & 4435.5\\ \bottomrule
   \end{tabular}
    \caption{Irradiation steps and corresponding total dose absorbed by the crystals}
    \label{tab:Irrstep}
\end{table}

\begin{figure}[h!]
    \centering
     \begin{tabular}{cc} 
    \includegraphics[width=0.49\textwidth]{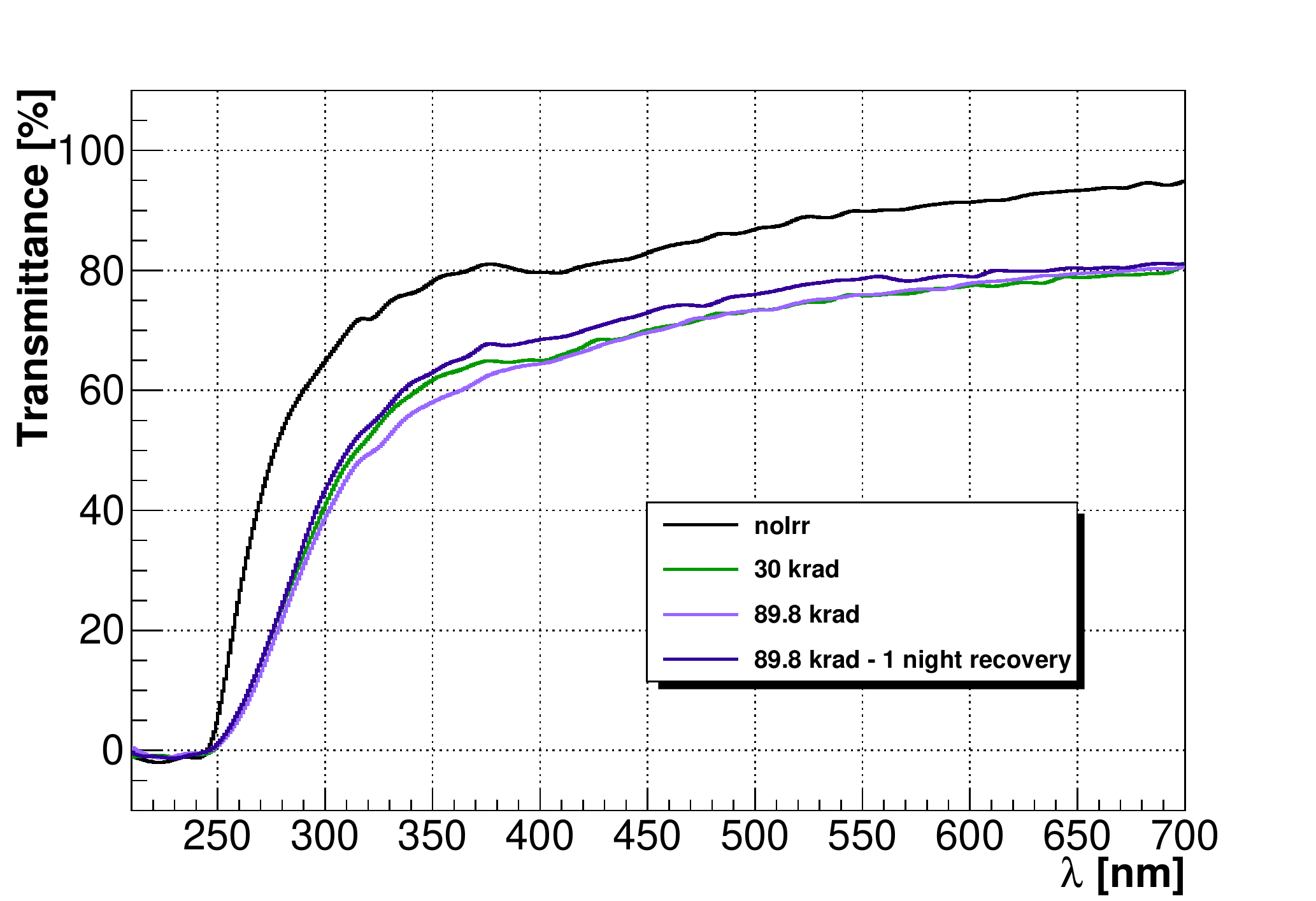} &\includegraphics[width=0.49 \textwidth]{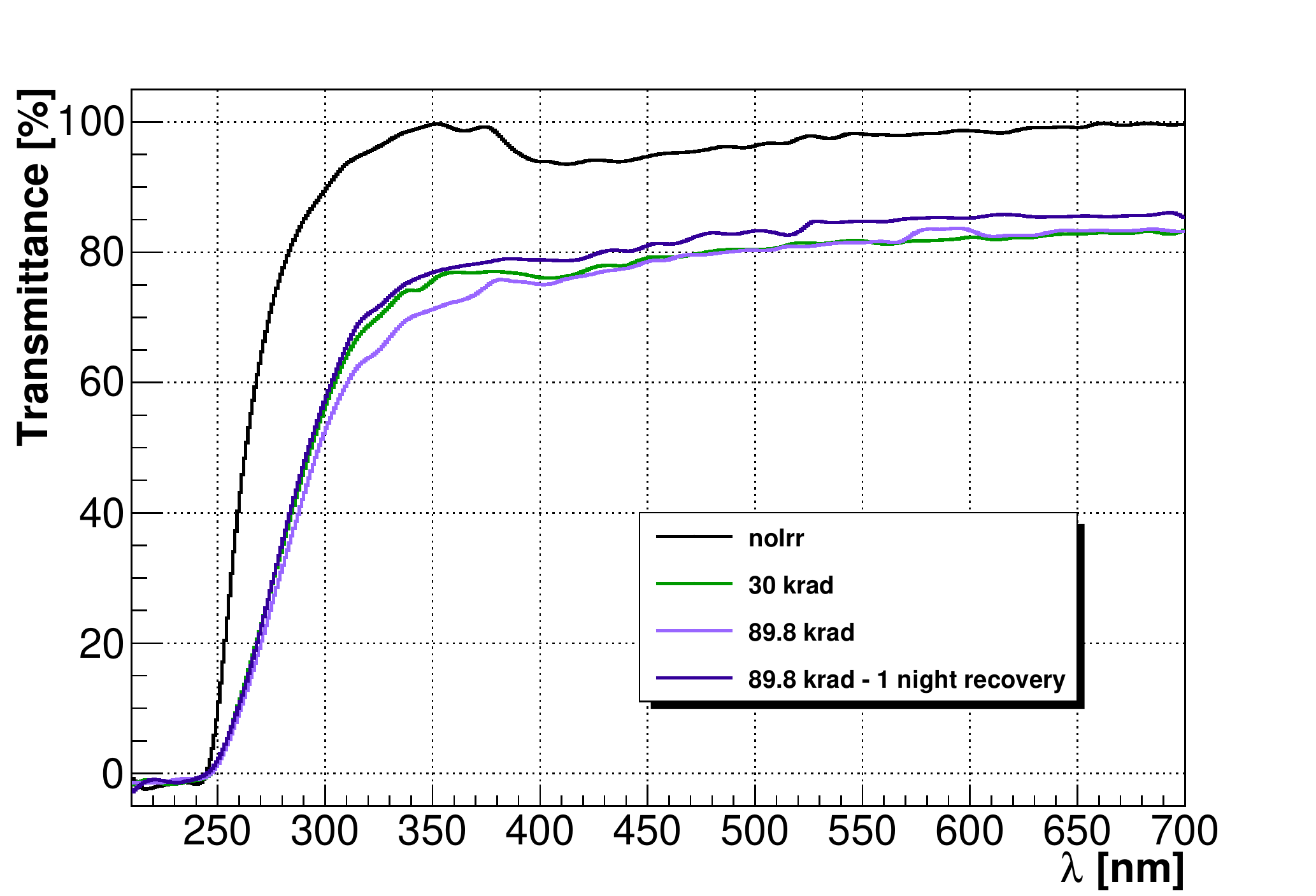} 
    \end{tabular}
    \caption{Transmission spectra obtained in step I and II for the naked crystal (left) and the crystal with Mylar wrapping (right)}
    \label{fig:trasm_s_I_II}
\end{figure}

In \figurename~\ref{fig:trasm_s_I_II} the spectra obtained for the first two irradiation steps are reported, as well as the spectra measured after keeping the crystals one night in a dark environment. A transmission recovery of ~10\% is observed at 350 nm. 

\begin{figure}[h!]
    \centering
     \begin{tabular}{cc} 
    \includegraphics[width=0.49 \textwidth]{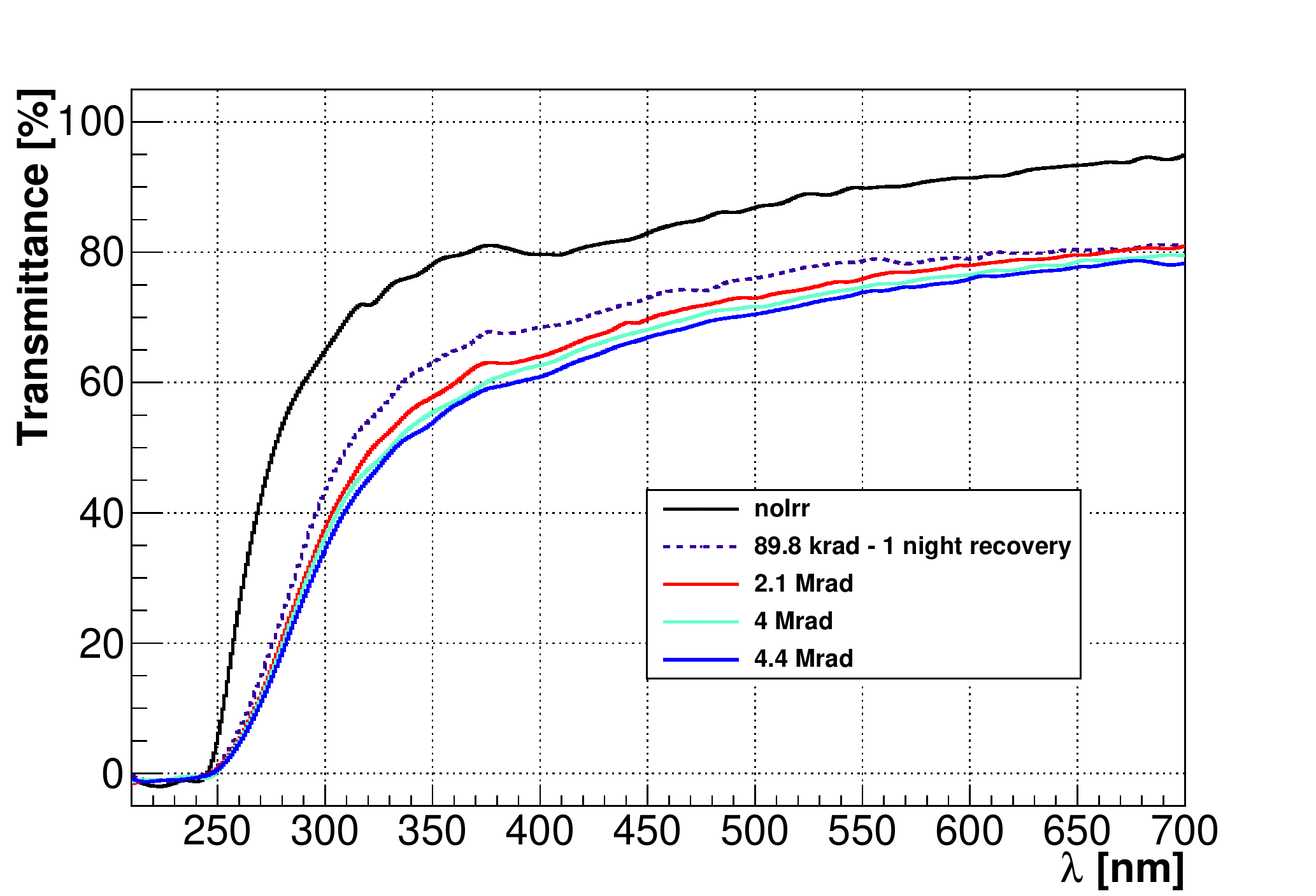} &\includegraphics[width=0.49 \textwidth]{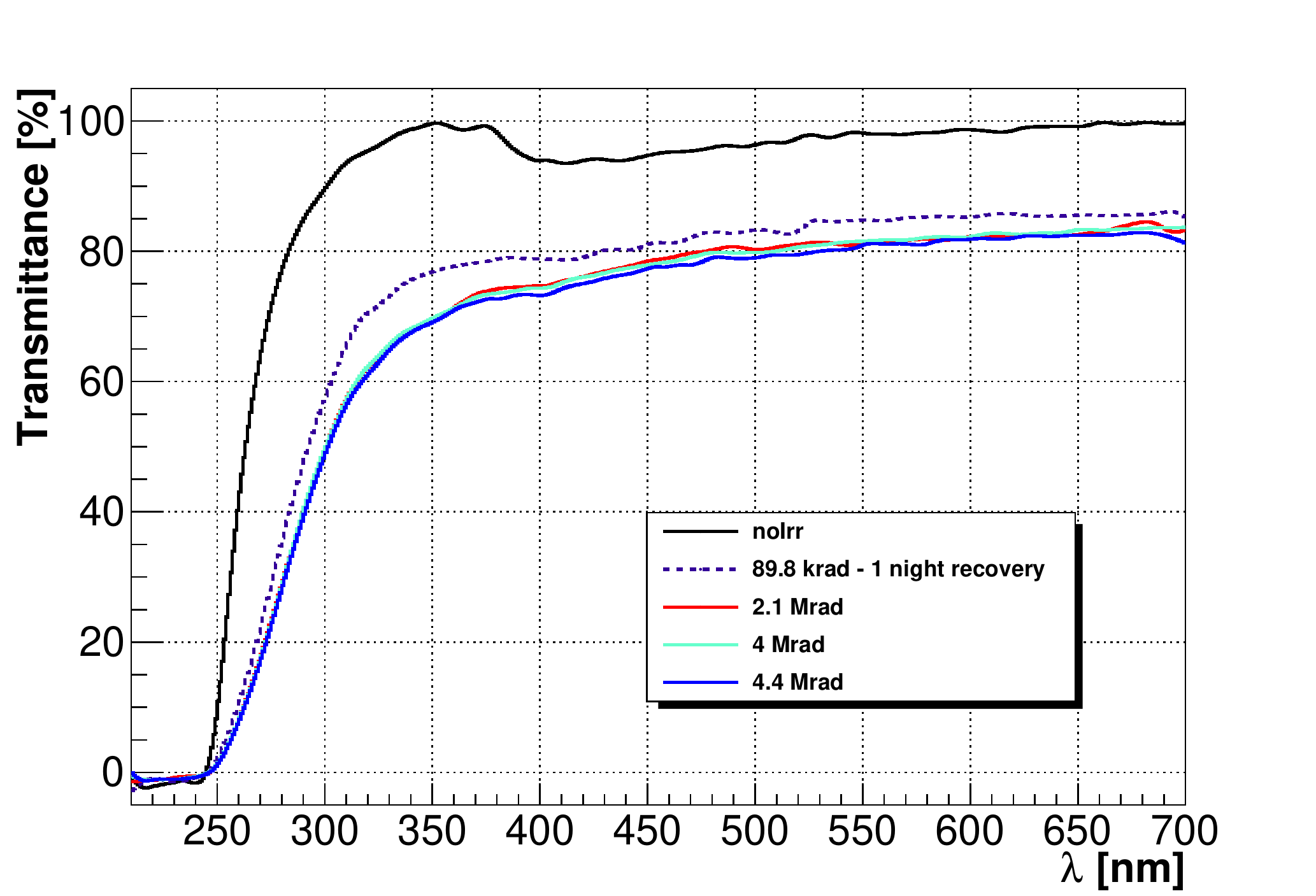} 
    \end{tabular}
    \caption{Transmission spectra obtained in step III, IV, V of the irradiation for the naked crystal (left) and the crystal with Mylar wrapping (right).}
    \label{fig:trasm_s_III_IV_V}
\end{figure}

\figurename~\ref{fig:trasm_s_III_IV_V} shows the longitudinal transmittance spectra obtained at step III, IV and V of the irradiation.

After a TID of approximately 80 krad, any subsequent irradiation step up to 4.4 Mrad did not result in any significant further decrease in transmittance, suggesting a saturation effect associated with the damage mechanism, as reported elsewhere \cite{zhu1996study}.
The maximum degradation observed is at the level of $\sim 40 \%$.
The results found with this study are within the range of degradation observed in~\cite{ren2001optical},~\cite{kozma2002radiation} and ~\cite{achenbach1998radiation}.

 \section{Natural annealing and optical bleaching of the crystals}
\begin{figure}[h!]
    \centering
     \begin{tabular}{cc} 
    \includegraphics[width=0.46 \textwidth]{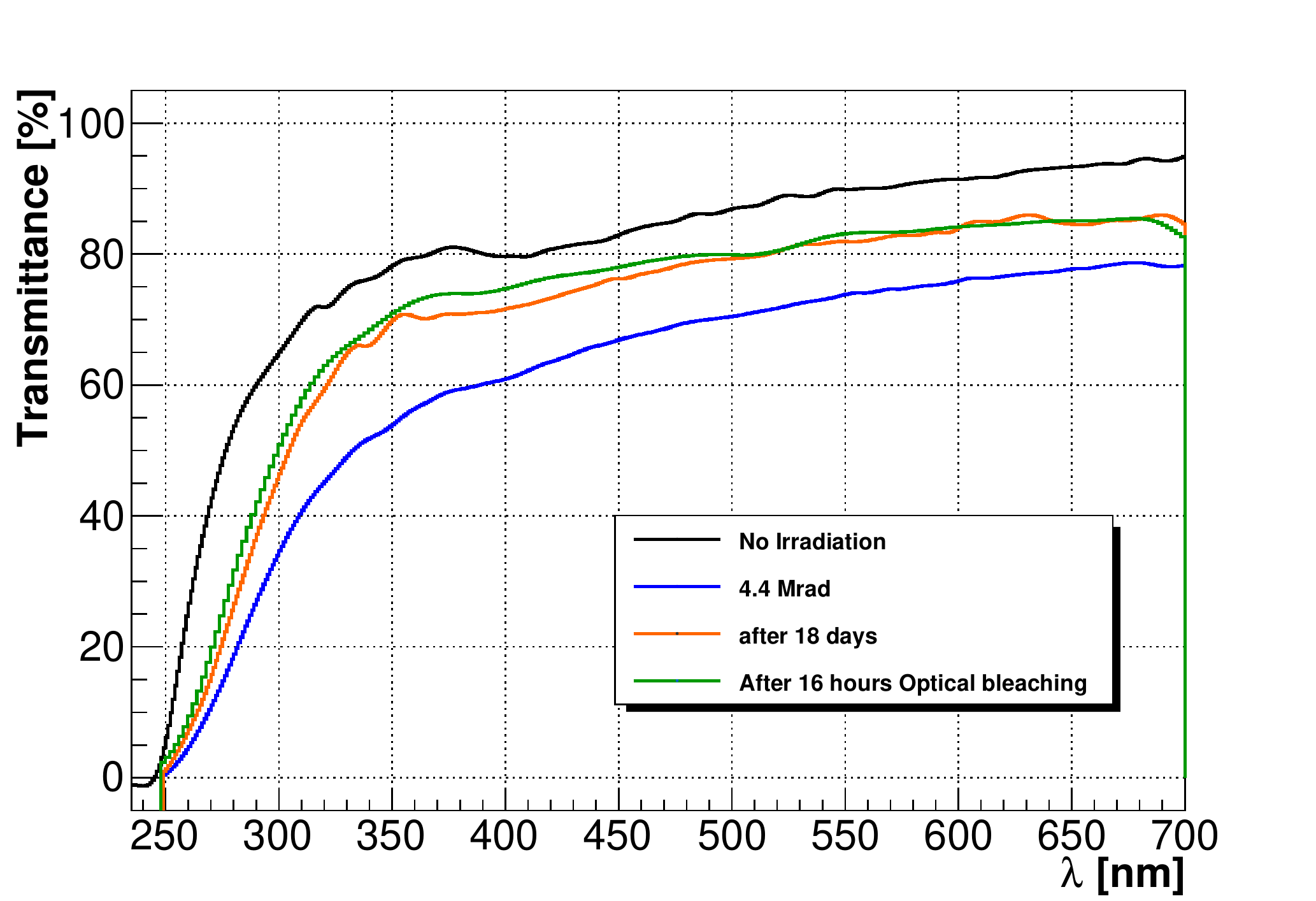} &
    \includegraphics[width=0.53 \textwidth]{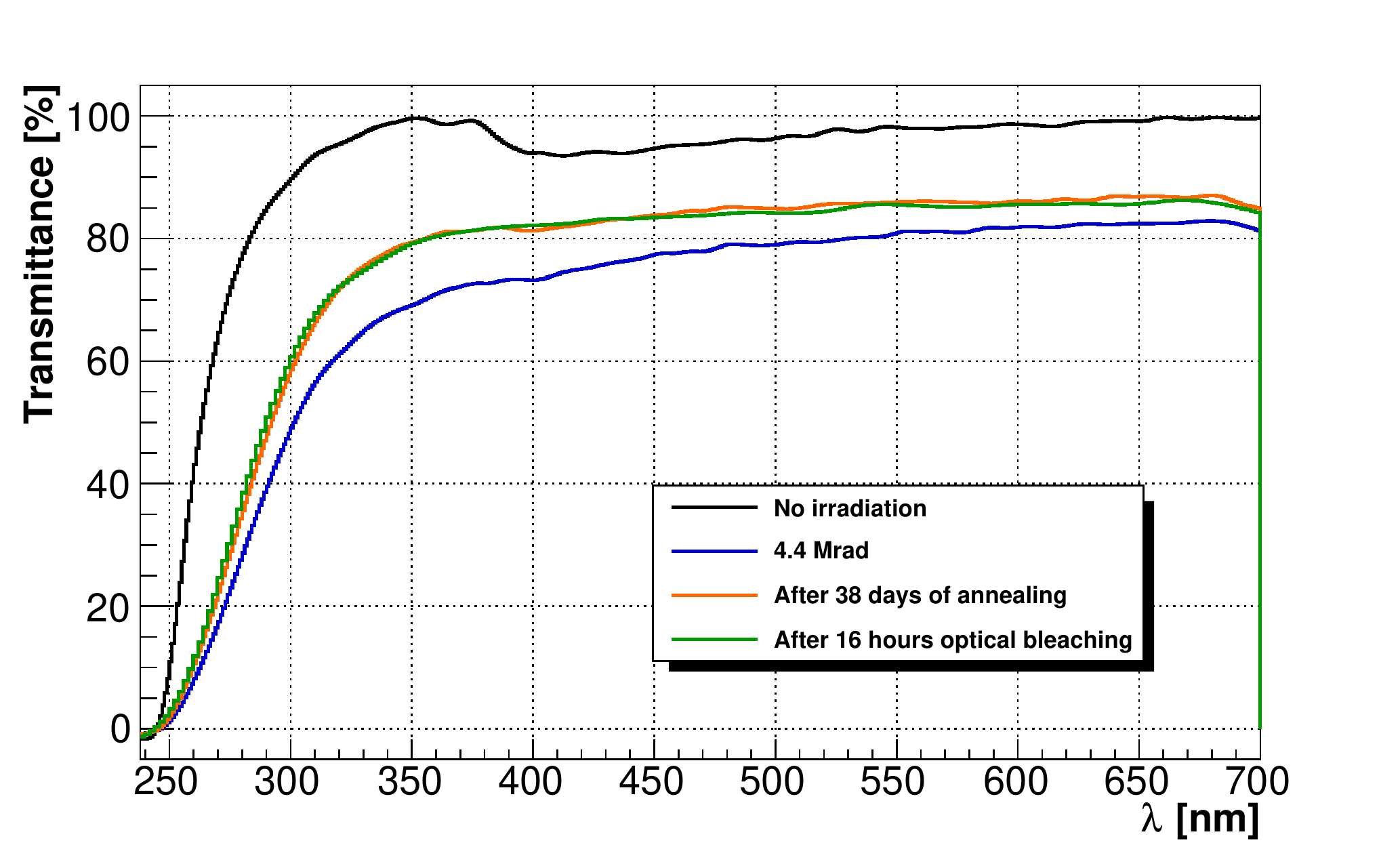} 
    \end{tabular}
    \caption{Transmission spectra obtained after 38 days of natural annealing and 16 hours of optical bleaching for the naked crystal (left) and for the crystal with Mylar wrapping (right).}
    \label{fig:ann_bl}
\end{figure}

 After the irradiation phase, crystals were kept in a dark box to allow their natural recovery, according to the procedure described in~\cite{kozma2002radiation}. To further improve the recovery process, a light-bleaching run with 400 nm blue light was performed on the irradiation crystals, according to the procedures and methods described in~\cite{achenbach1998radiation}.
 After 16 hours of optical bleaching, a recovery of a few percent was observed in the transmittance spectrum, indicating a small accelerating effect on the recovery process with respect to the natural annealing case.
 A summary of the results obtained with natural annealing and the optical bleaching can be found in \figurename~\ref{fig:ann_bl}.

\section{Optical properties after neutron irradiation}
\begin{figure}[h!]
    \centering
    \includegraphics[width=0.49\textwidth]{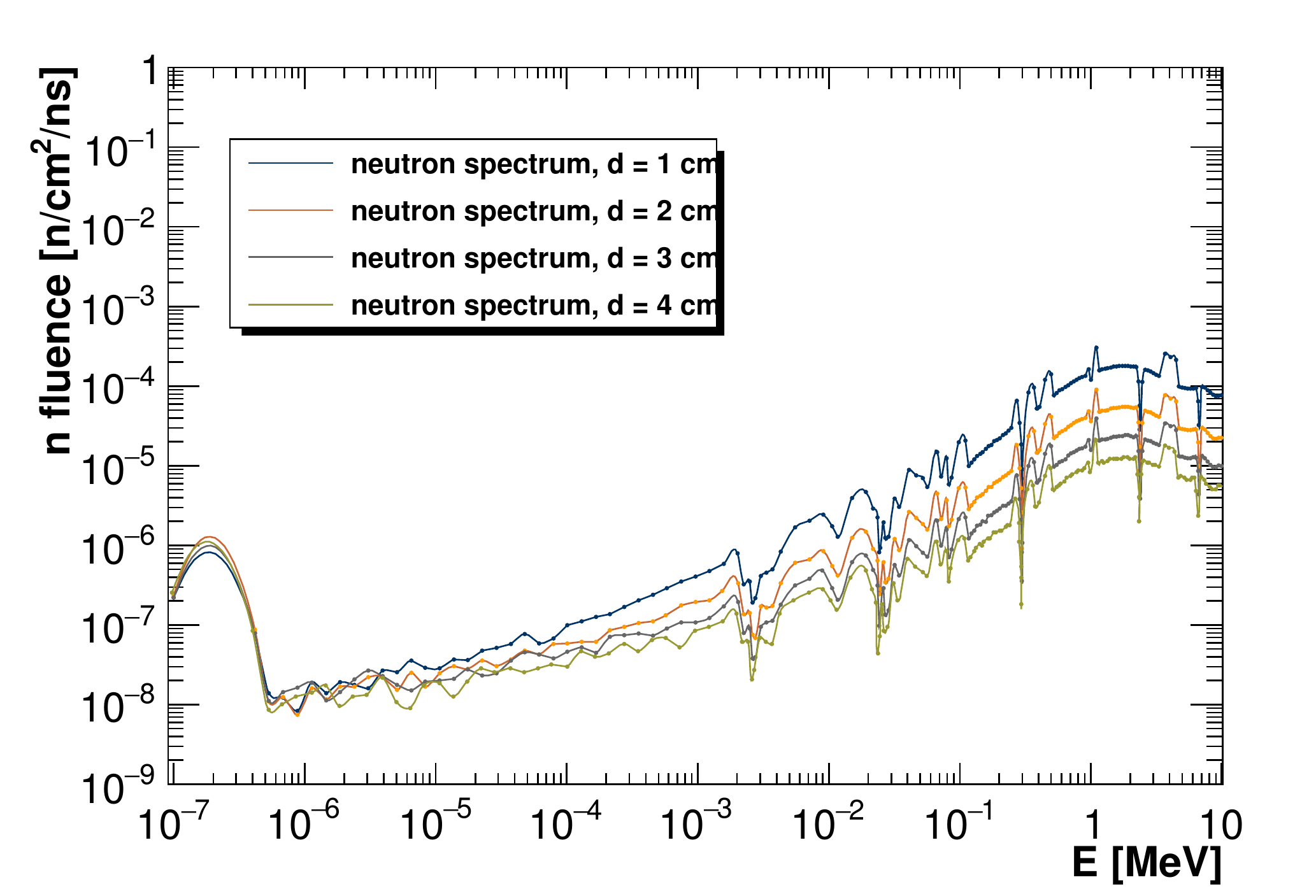}
    \caption{FNG neutron spectra as a function of the fluence.}
    \label{fig:NN}
\end{figure}

The crystals were subsequently irradiated at the Frascati Neutron Generator (FNG) facility of ENEA Frascati \cite{FNG}.\\
Neutron generation at FNG is based on the T(d,n)$\alpha$ fusion reaction, producing 14 MeV neutrons with a flux up to ~$10^{12}$ neutrons/s in steady state or pulsed mode. In Figure~\ref{fig:NN}, the neutron spectrum is shown as a function of the fluence.

\begin{figure}[h!]
    \centering
    \begin{tabular}{cc}
 \includegraphics[width=0.45 \textwidth]{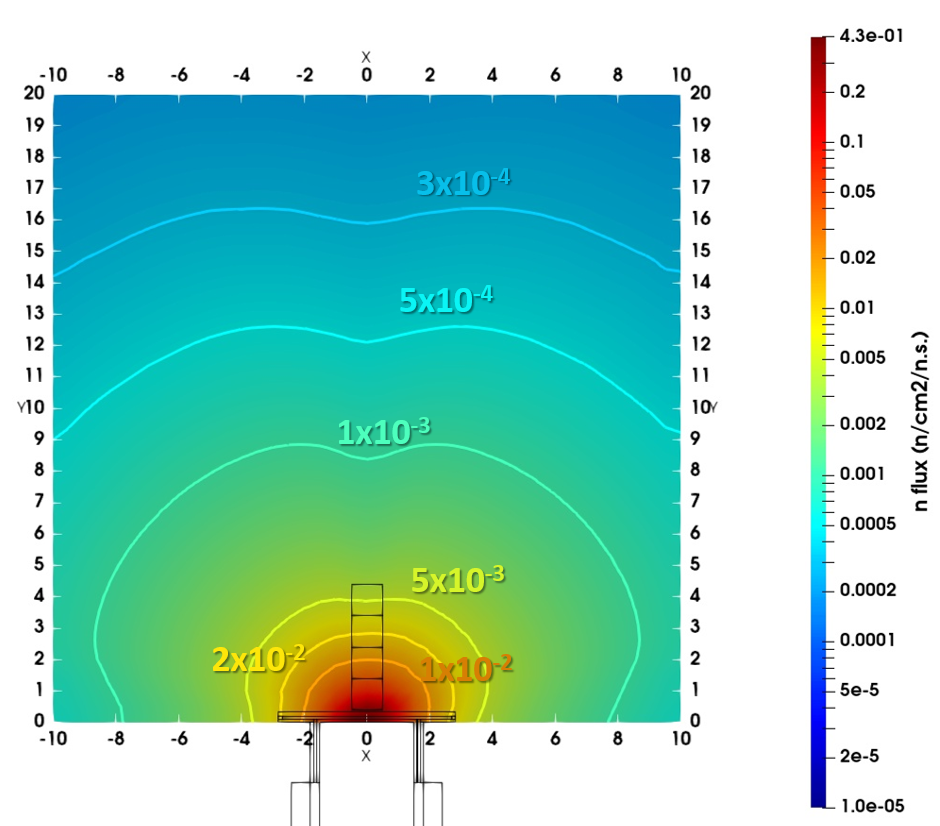}&  
 \includegraphics[width=0.42 \textwidth]{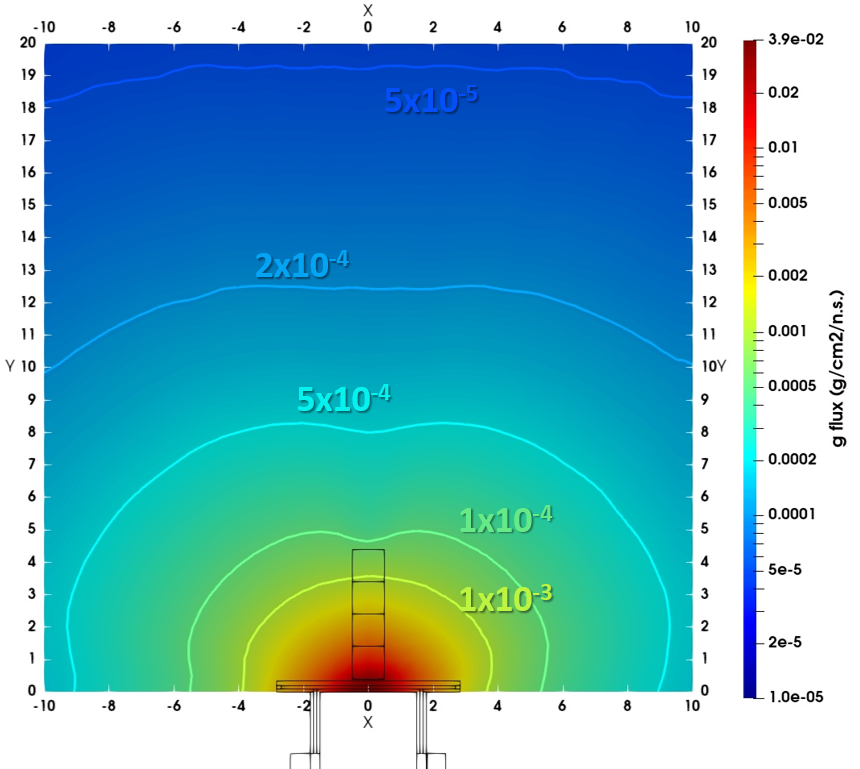}
    \end{tabular}
    \caption{Simulation of the neutron (left) and gamma (right) flux: crystals were placed 1 cm away from the neutron gun.}
    \label{fig:XX}
\end{figure}

The crystals under test were placed 1 cm away from the source (relatively to their front face).
Figure~\ref{fig:XX} (left) reports the neutron flux map in the FNG facility. In order to better evaluate the expected neutron fluences, a simulation of the irradiation process was implemented on the McStas simulation package \cite{mcstas}, using 1 cm bins along the crystal axes. \\
As shown in Figure \ref{fig:XX} (right), the TID associated with the neutron irradiation process resulted in a (negligible) 0.1 krad dose for crystals.\\

\begin{figure}[h!]
    \centering
     \begin{tabular}{cc} 
    \includegraphics[width=0.49 \textwidth]{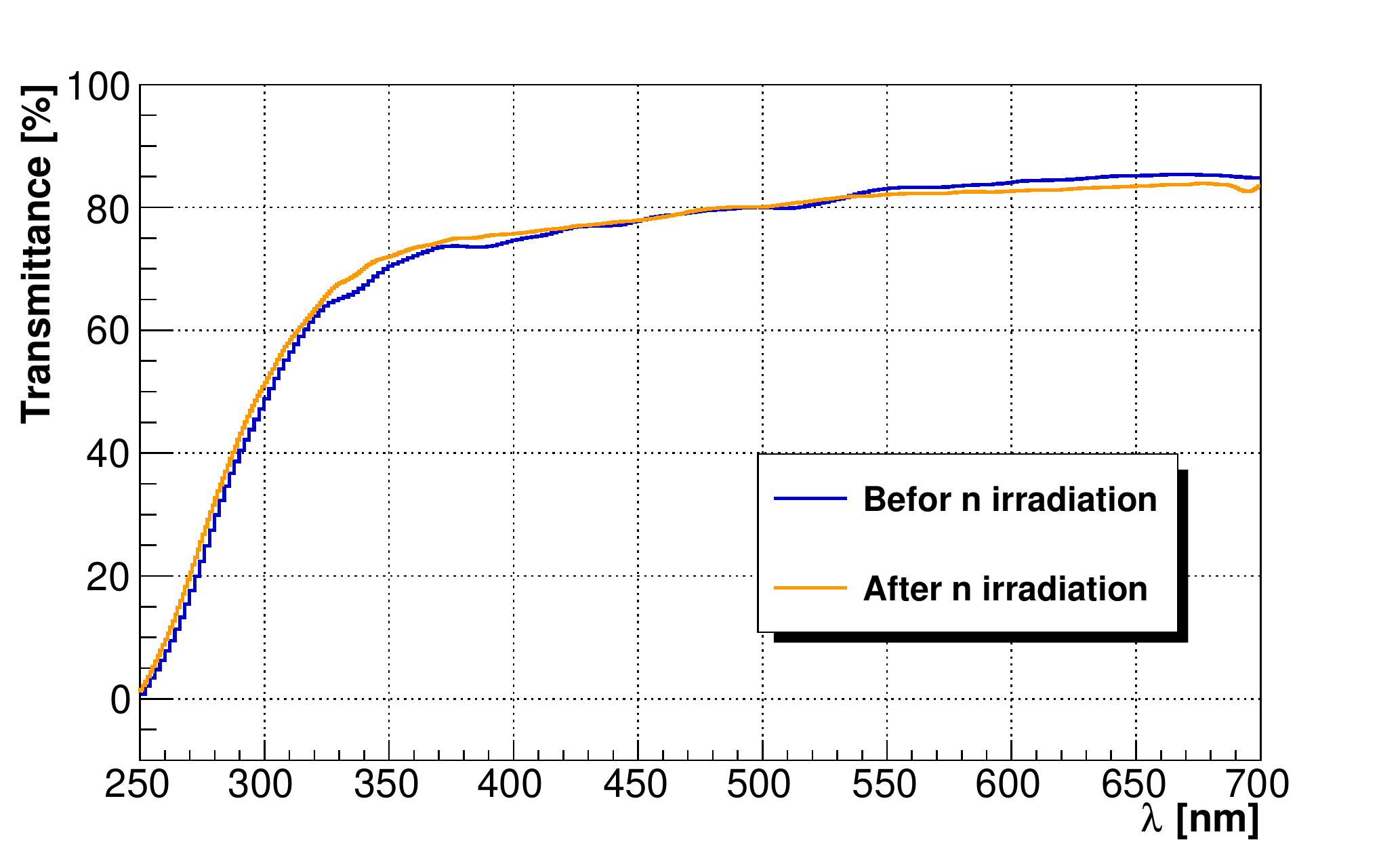} & 
    \includegraphics[width=0.49 \textwidth]{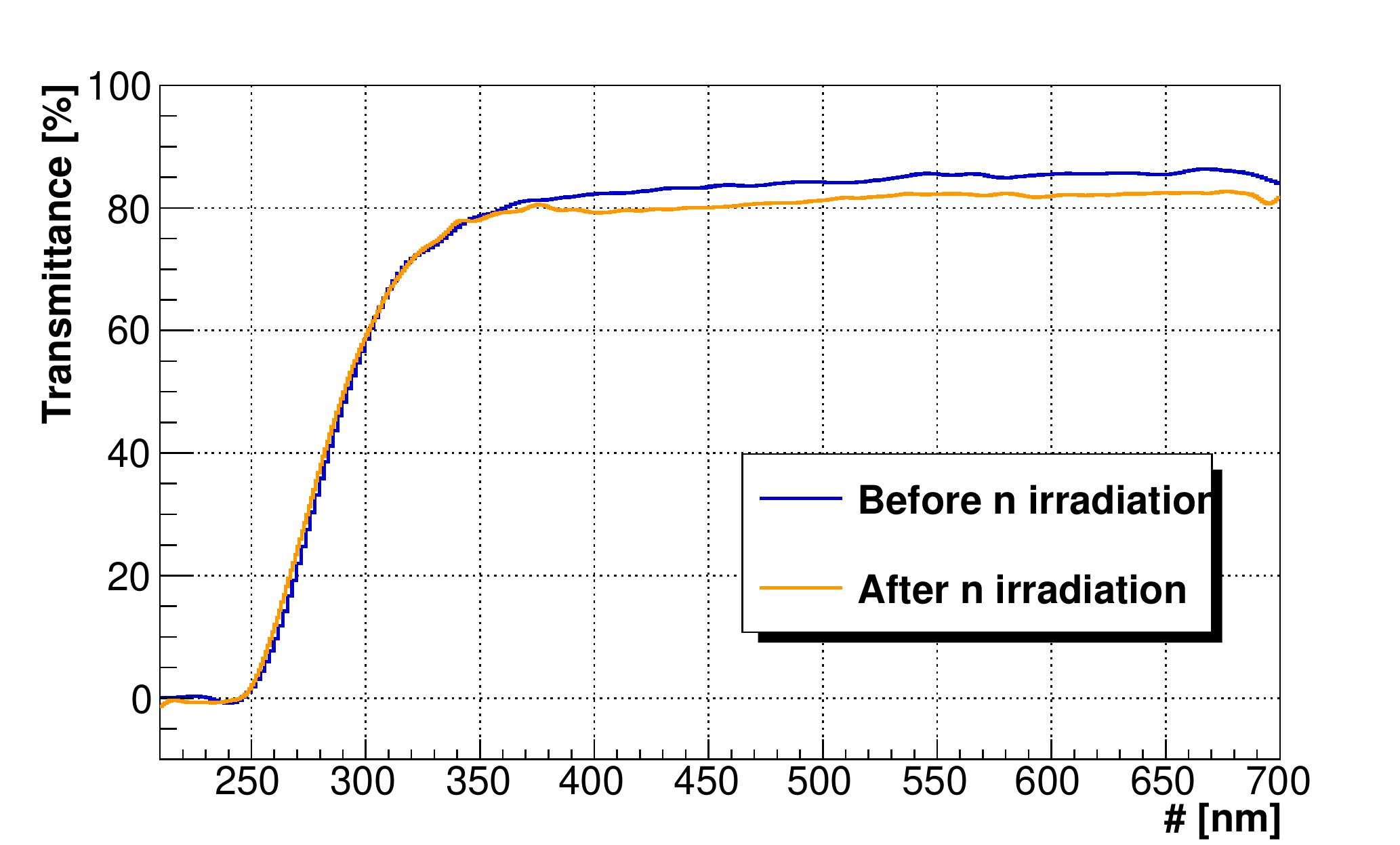} 
    \end{tabular}
    \caption{Transmission spectra obtained after the  irradiation at FNG with 14 MeV neutrons for a total fluence of $10^{13}$n/cm$^2$ (orange line) compared with the results  after the 16 hours optical bleaching (blue line) for the naked crystal (left) and for the crystal with Mylar wrapping (right).}
    \label{fig:DopoNeutroni}
\end{figure}

The total fluence delivered in the first cm of the crystals during $\sim 1$ hour and 30 minutes of irradiation with 14 MeV neutrons was $10^{13}$ n/cm$^2$.\\
Due to the technical time related to logistics and shipment of the crystals, transmittance measurements could only be performed 14 days after the irradiation and showed no alteration in the transmittance spectrum (\figurename~\ref{fig:DopoNeutroni} ).

\newpage
\section{Conclusion}
The optical transmittance of the two SICCAS \pbfd crystals, the first without any wrapping and the second one with a Mylar wrapping, was studied before and after irradiation performed with photons and neutrons.\\
The transmittance of the crystals under test were monitored during the photon irradiation steps, up to a 4.4 Mrad TID,  and after the subsequent neutron irradiation runs. Natural recovery of crystals, along with light-bleaching and thermal annealing were carried out and compared in terms of their effect on the recovery process.

\acknowledgments

This work was developed within the framework of the International Muon Collider Collaboration (\url{https://muoncollider.web.cern.ch}), where the Physics and Detector Group aims to evaluate potential detector R\&D to optimize experiment design in the multi-TeV energy regime.


\bibliography{biblio}
\end{document}